\documentclass[conference]{IEEEtran}
\IEEEoverridecommandlockouts

\usepackage{cite}
\usepackage{url} 
\usepackage{amsmath,amssymb,amsfonts}
\usepackage{algorithmic}
\usepackage{graphicx}
\usepackage{textcomp}
\usepackage{xcolor}

\def\BibTeX{{\rm B\kern-.05em{\sc i\kern-.025em b}\kern-.08em
    T\kern-.1667em\lower.7ex\hbox{E}\kern-.125emX}}
\begin{document}

\title{IA-MoE: An Intensity-Aware Mixture of Experts for DNN Inference
}

\author{\IEEEauthorblockN{Meiyan Zeng, Sasu Tarkoma, Xiaoli Liu}
\IEEEauthorblockA{\textit{University of Helsinki, Finland. Email: \{firstname.lastname\}@helsinki.fi} 
}
}

\maketitle

\begingroup
\footnotesize
\noindent This work extends our previous paper published at IEEE ICNSC 2025.\par
\noindent\textbf{Conference paper DOI:} \url{https://doi.org/10.1109/ICNSC66229.2025.00081}\par
\par\medskip
\endgroup

\begin{abstract}
     In edge-cloud collaborative inference, the layers of a deep neural network (DNN) are placed on edge and cloud virtual machines (VMs) for every request, while the edge load and the cross-cloud bandwidth change over time. We propose IA-MoE, an intensity-aware Mixture of Experts (MoE) framework that scores each layer-VM pair by end-to-end delay and energy and places the layers one at a time on the best-scoring VM. Such greedy placement is myopic when trained on per-layer labels: offloading to the cloud slows the first offloaded layer while the gain appears only in later layers, so the policy keeps too many requests at the edge. Training on the remaining request time after each decision removes this myopia and reduces the mean delay by 12--56\% on congested workloads. In CloudSimSDN simulations with three DNNs and two public request traces, IA-MoE has the lowest mean delay of all compared methods on congested VGG16 workloads, across cross-cloud bandwidths of 150--600~Mbps, fluctuating bandwidth and saturated edge servers. A random forest trained on the same label is 5--27\% slower, and only a heuristic that sends whole requests to the cloud uses less energy, at up to 20 times the delay. When the edge is lightly loaded, offloading does not pay and simple heuristics perform as well. IA-MoE also beats a Double Deep Q-Network (DQN) agent with the same features in all 19 tested settings, with a mean delay lower by a factor of 1.3 to 10.5, and trains in minutes instead of hours.
\end{abstract}


\section{Introduction}
In recent years, artificial intelligence (AI) technologies, particularly deep neural networks (DNN), have achieved significant progress in various application fields, such as autonomous driving \cite{ma2020artificial, ling2022blastnet} and medical image recognition \cite{liu2021advances, zhou2023volumetric}. However, as parameter scale and complexity grow, traditional cloud-based AI inference services are increasingly facing bottlenecks in performance, privacy protection, and energy consumption management. Uploading raw data from end devices to the cloud for inference causes significant network delays and sensitive data leakage risks \cite{zhou2019edge}. Furthermore, high energy consumption in data centers has hindered large-scale AI deployment. A power analysis report \cite{powermagwebsite} shows that global data centers consumed 340 TWh annually in 2023, accounting for 1.3\% of global electricity use.

Edge computing (EC) and edge intelligence (EI) address these issues by offloading inference to edge servers near the data source \cite{zhou2019edge}. This approach reduces latency and enhances privacy protection while leveraging the hybrid computing resources of edge and cloud. In response to the high computational demands of DNN models during the inference stage, researchers \cite{kang2017neurosurgeon, li2018edge} have proposed model partitioning and collaborative inference, decomposing DNNs into hierarchical structures of several independent layers (or blocks) and offloading partial layers to different computing nodes for execution. By leveraging the much smaller size of intermediate-layer features compared to raw inputs, appropriate model partitioning and task offloading can reduce data transmission and end-to-end latency, thereby enhancing inference efficiency and energy performance.

Partitioning decisions in a hybrid edge-cloud system have to be made for every request, while request rates and input sizes, the load on the edge VMs and the cross-cloud bandwidth change over time. Most existing schemes consider a single end device or a single edge server \cite{chen2021dnnoff, karjee2022split, yan2022optimal}, leave out the cloud \cite{karjee2022split, yan2022optimal, lin2024murmuration, taufique2024hidp}, or optimize latency alone \cite{lin2019cost, chen2021dnnoff}. With several edge servers, a cloud and two objectives, the placement problem is typically NP-hard \cite{lim2024cutting}, so exact solvers are too slow for per-request decisions. A learned model that scores each pair of a layer and a VM is fast enough, and the layers of a request can then be placed one at a time on the VM with the highest score. This greedy placement, however, does not see the full cost of a decision at the layer where it is made. Moving a request to the cloud makes the first offloaded layer slower, because its input must cross the wide-area link, while the gain appears in the later layers. A model trained to score each layer by its own cost therefore offloads far too little, even when offloading is better for the whole request. Greedy layer-by-layer offloading is known to be suboptimal when the layer costs are given \cite{lin2019cost}; we show that the same myopia appears in a learned score.

In this paper, we propose IA-MoE, an intensity-aware Mixture of Experts (MoE) framework that dynamically determines the partitioning and scheduling of DNN inference tasks by matching each layer task to a VM in hybrid edge-cloud computing environments. We formulate the placement as a mixed-integer program (MIP) that minimizes a weighted sum of end-to-end latency and energy consumption. To remove the myopia, we train IA-MoE on a label that measures the remaining time of the request after the decision; the placement procedure stays the same, and training on this label approximates a one-step policy improvement over the data-collection policies \cite{brandfonbrener2021offline}. Earlier learning-based offloading imitates optimal decisions found offline by exhaustive search or a solver, one application at a time \cite{yu2020intelligent, chen2020knowledge}; with requests queueing for the same VMs and links and a fluctuating cross-cloud bandwidth, such optimal decisions are not available, and our label is computed from the logged outcomes of simple collection policies. The experts are organized by the operational intensity of the VMs.

The main contributions of this paper are as follows:

\begin{itemize}
    \item We identify the myopia of greedy layer-by-layer placement trained on per-layer labels and remove it with a remaining-time label, which approximates a one-step policy improvement over the data-collection policies \cite{brandfonbrener2021offline} and reduces the mean delay by 12--56\% on congested workloads.
    \item We show that this supervised matching beats a Double Deep Q-Network (DQN) agent \cite{vanhasselt2016deep} with the same features, network size and simulator in all 19 tested settings, while training in minutes instead of hours; unlike earlier imitation learning for offloading \cite{yu2020intelligent, chen2020knowledge}, it needs no optimal decisions as labels.
    \item  In CloudSimSDN simulations with three DNN models (AlexNet, VGG16 and ResNet50) and two public request traces, IA-MoE has the lowest mean delay of all compared methods on congested VGG16 workloads, for mean cross-cloud bandwidths of 150--600~Mbps, a coefficient of variation (CV) of the bandwidth up to 0.5 and saturated edge servers, and performs on par with simple heuristics on light workloads. A Roofline bound shows that dividing experts by operational intensity can help only when some devices have more compute and others more memory bandwidth; ablations confirm that it has no measurable effect in our configuration, which lacks such devices. The code is available online.\footnote{\url{https://github.com/mayrazing/IA-MoE}}
\end{itemize}

\section{Problem Formulation}\label{sec_problem}
\begin{figure}[htbp]
\centerline{\includegraphics[width=\linewidth]{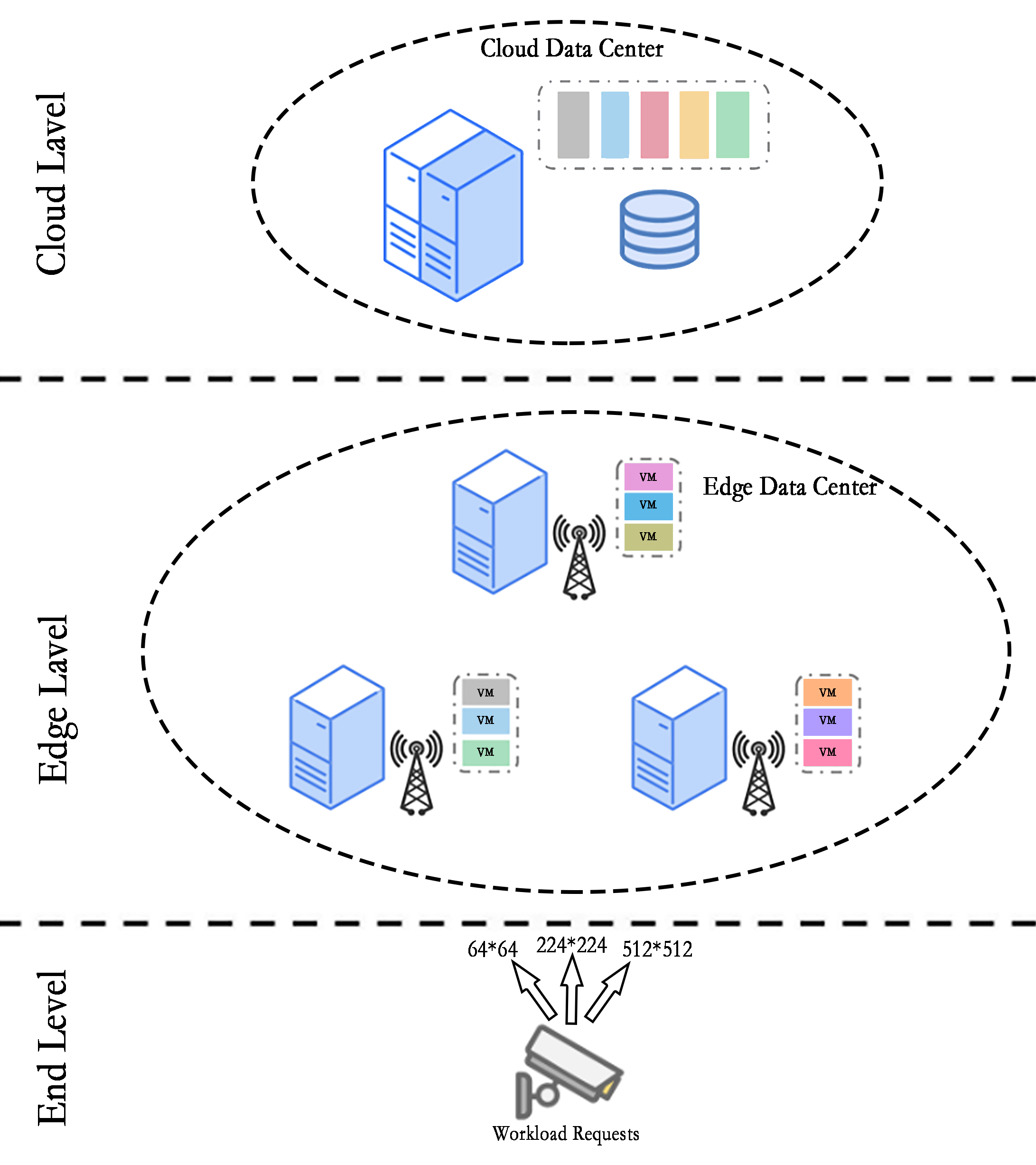}}
\caption{Three-tier Structure.}
\label{fig_system_model}
\end{figure}

The proposed architecture follows a three-tier hybrid design consisting of end, edge, and cloud layers shown in Fig.~\ref{fig_system_model}. The end layer comprises end devices hosting the DNN model and generating a stream of inference requests such as facial recognition images. To improve utilization and scalability, servers are virtualized into heterogeneous virtual machines (VMs)/containers, enabling concurrent service for multiple requests. Given the memory-sharing capability of VM-based data centers, all pre-trained DNN models are assumed to reside in the shared memory pool.

\subsection{System Environment}
We use a set \( C = \{c_1, \dots, c_I\} \) to represent all computing entities in the three-tier computing architecture, including cloud servers, edge servers, and end devices. Entity $i$ is equipped with a CPU having \( c_i^{\text{core}} \) homogeneous cores and the computing power per core \( c_i^{\text{pflops}} \). The memory bandwidth and network bandwidth of the entity are \( c_i^{\text{ramBw}} \) and \( c_i^{\text{bw}} \) respectively. For simplicity, we also use notations \( C_{\text{cloud}} \), \( C_{\text{edge}} \) and \( C_{\text{end}} \) to distinguish cloud, edge, and terminal entities respectively.

Similarly, we use \( \mathbb{M} = \{m_1, \dots, m_R\} \) to represent the set of VM instances placed on any \( c_i \in C \), where $ m_r = \langle m_r^{\text{core}}, m_r^{\text{pflops}}, m_r^{\text{ramBw}}, m_r^{\text{bw}} \rangle $ represents the vCPU cores, per-core computing rate, memory bandwidth, and network bandwidth allocated to the r-th VM. To ensure uniqueness, each VM can only be placed on one computing entity:

\begin{equation}
\begin{split}
\sum_{c_i \in C} y_{c_i}^{m_r} = 1, \quad \forall m_r \in \mathbb{M},
\end{split}
\label{eq_1}
\end{equation}

where if VM \(m_r\) is placed on node \(c_i\), then the binary variable $y_{c_i}^{m_r} = 1$, otherwise 0.

\subsection{DNN Partitioning}
 The DNN model is represented as a directed acyclic graph (DAG) \(G = (L, E)\), where $L$ is the set of inference layers and $E$ is the set of dependency edges. The set $L_W \subseteq L$ contains the \(|L_W|\) non-virtual layers: \(l_1, l_2, \dots, l_{|L_W|}\), and an edge \((l_x, l_y)\) means \(l_x\) must be computed before \(l_y\), with \(l_x\) as the predecessor and \(l_y\) as the successor. For layer \(l_k\), \(\text{pre}(l_k)\) and \(\text{suc}(l_k)\) denote its predecessor and successor sets, respectively. Thus, the input data of \(l_k\) is the output data of \(\text{pre}(l_k)\), and \(l_k\) can only start after all predecessors are completed.

We focus on a set $R$ of DNN inference requests from the same end device. For each request \(r_w \in R\), the final output must be transmitted back to the end device once all layers complete execution. Following the approach in \cite{li2021throughput}, two virtual layers are added to the DNN's DAG: \(l_0\) (input data layer) and \(l_{|L_W|+1}\) (output result layer), so that $L = L_W \cup \{l_0, l_{|L_W|+1}\}$. The output size of \(l_0\) is the original input data size, and the input size of \(l_{|L_W|+1}\) is the result size to be transmitted back. The DNN partitioning and placement model introduces two sets of binary  decision variables: \(y_{m_r}\), which equals 1 if VM \(m_r\) is selected for partitioning (0 otherwise), and  \(b_{m_r}^{l_k}\), which equals 1 if layer \(l_k\) is allocated to VM \(m_r\) (0 otherwise). The following constraints are then defined: 

\textit{1) Only one end device serves the DNN model:} 

\begin{equation}
\begin{split}
\sum_{c_i \in C_{\text{end}}} \sum_{m_r \in \mathbb{M}} y_{m_r} \cdot y_{c_i}^{m_r} = 1.
\end{split}
\end{equation}

\textit{2) Each non-virtual layer is served by exactly one VM: } 

\begin{equation}
\begin{split}
\sum_{m_r \in \mathbb{M}} b_{m_r}^{l_k} = 1, \quad \forall l_k \in L_W.
\end{split}
\end{equation}

\textit{3) All layers must be allocated to selected VMs:}

\begin{equation}
\begin{split}
\sum_{l_k \in L_W} \sum_{m_r \in \mathbb{M}} b_{m_r}^{l_k}\cdot y_{m_r} = |L_W|.
\end{split}
\end{equation} 

\textit{4) Virtual layers \(l_0\) and \(l_{|L_W|+1}\) must be allocated to the same end device:}

\begin{equation}
\begin{split}
\sum_{c_i \in C_{\text{end}}} \sum_{m_r \in \mathbb{M}} b_{m_r}^{l_0} \cdot y_{c_i}^{m_r} = 1.
\end{split}
\end{equation}

\begin{equation}
\begin{split}
\sum_{c_i \in C_{\text{end}}} \sum_{m_r \in \mathbb{M}} b_{m_r}^{l_{|L_W|+1}} \cdot y_{c_i}^{m_r} = 1.
\end{split}
\label{eq_2}
\end{equation}

\subsection{End to End Delay}
For a request from end device \(c_i \in C_{\text{end}}\), the end to end (E2E) latency \(T_{\text{req}}\) consists of four parts:

\begin{equation}
\begin{split}
T_{\text{req}} = T_{\text{comm\_up}} + T_{\text{prep}} + T_{\text{total\_process}} + T_{\text{comm\_down}}.
\end{split}
\end{equation}

\textit{1) Transmission time during the terminal data upload stage:} Let \(D_{l_0}\) be the output size of virtual layer \(l_0\), and \(B_{c_i}^{\text{ul}}\) be the uplink bandwidth of \(c_i\). The transmission time of \(D_{l_0}\) (raw data input) is given by

\begin{equation}
\begin{split}
T_{\text{comm\_up}} = \frac{D_{l_0}}{B_{c_i}^{\text{ul}}}.
\end{split}
\end{equation}

\textit{2) Transmission time during the final result return stage:} Similar to the transmission time of raw data input, we can derive the transmission time of the final results back to the end device as

\begin{equation}
\begin{split}
T_{\text{comm\_down}} = \frac{D_{l_{|L_W|+1}}}{B_{c_i}^{\text{dl}}},
\end{split}
\end{equation}
where \(D_{l_{|L_W|+1}}\) denotes the result size of virtual layer \(l_{|L_W|+1}\), and \(B_{c_i}^{\text{dl}}\) denotes the downlink bandwidth of \(c_i\).

\textit{3) Execution time of the server preprocessing stage:} The time consumption of preprocessing on the server depends on sample size and the computational resources. Since it is executed only once and its computational complexity is much lower than that of the inference phase, it represents a lightweight task for edge/cloud servers. To simplify the E2E delay problem and focus on DNN partitioning strategies, \(T_{prep}\) is modeled as a fixed item (such as 0.001s) similar to VM event scheduling overhead, which does not cause a substantial deviation to the final layer partitioning or scheduling results.

\textit{4) Processing time of the server inference stage:} For the subtask \(l_k\) running on each VM instance \(m_r\), the execution delay \(T_{\text{exec}}(l_k)\) can be described by the following equation:

\begin{equation}
\begin{split}
T_{\text{exec}}(l_k) = T_{\text{process}}(l_k) + T_{\text{trans}}(l_k).
\end{split}
\end{equation}

Here, processing latency \(T_{\text{process}}(l_k)\) includes memory access and computation latency, which can be given by

\begin{equation}
\begin{split}
T_{\text{process}}(l_k) = T_{\text{memory}}(l_k) + T_{\text{comp}}(l_k) \\ =  \frac{A_{l_k}}{B_{\text{mb}}} + \frac{X_{l_k}}{m_r^{\text{core}} * m_r^{\text{pflops}}},
\end{split}
\end{equation}
where \(A_{l_k}\) is the memory access volume, \(B_{\text{mb}}\) is the VM's memory bandwidth, and \(X_{l_k}\) is the computation volume, counted in multiply-accumulate operations (MACs); accordingly, $m_r^{\text{pflops}}$ is the per-core rate in MACs per second. The two terms are added rather than overlapped, so memory bandwidth affects every layer.

The cross-node transmission delay, \(T_{\text{trans}}(l_k)\), accounts for the time required to transmit the output feature map (i.e., intermediate activation values) and is defined as:

\begin{equation}
\begin{split}
T_{\text{trans}}(l_k) = \frac{D_{l_k}}{m_r^{\text{bw}}},
\end{split}
\end{equation}
where \(D_{l_k}\) represents the intermediate output data size of subtask \(l_k\) ( layer \(l_k\)). Here $m_r^{\text{bw}}$ stands for the bottleneck bandwidth of the path; on paths between the edge and the cloud it is the cross-cloud bandwidth, which fluctuates over time (Section~\ref{sec_settings}). The formulas give the delay without contention; the simulator additionally models queueing on the VMs and on the links, which is the main source of delay under load.

Additionally, we assume that the arrival time \(\text{Arrival}(l_k)\) of subtask \(l_k\), representing the latest time when all direct predecessor layers of this layer have been processed, is given by 

\begin{equation}
\begin{split}
\text{Arrival}(l_k) = \max_{l_j \in \text{pre}(l_k)} \left( \text{Arrival}(l_j) + T_{\text{exec}}(l_j) \right),
\end{split}
\end{equation}
where \(\text{pre}(l_k)\) denotes the set of the direct predecessor layers of layer \(l_k\).

In summary, the server processing stage \(T_{\text{total\_process}}\) for each inference request \(r_w\) can be expressed as the arrival time of the virtual subtask \(l_{|L_{W}| + 1}\) plus the processing delay of this layer.

\begin{equation}
\begin{split}
T_{\text{total\_process}} = \text{Arrival}(l_{|L_{W}| + 1}) + T_{\text{process}}(l_{|L_{W}| + 1}).
\end{split}
\end{equation}

\subsection{Energy Consumption}\label{sec_settings_energy}
We consider the computational energy of the layers and the energy of data transmission. When server \(c_i\) is in use, the static power is \(P_{\text{idle}}(c_i)\), and the operating power at full load is \(P_{\text{max}}(c_i)\). Thus, the operating power under the actual load is $P_{\text{run}}(c_i) = P_{\text{max}}(c_i) \cdot \text{CPU\_Util}(c_i)$, where \(\text{CPU\_Util}(c_i)\) represents the CPU utilization rate of this server. When allocating the total power of the server to its VM instances, we scale it proportionally. Specifically, for each VM instance \(m_r\), the fixed power consumption of \(m_r\) is \(P_{\text{idle}}(m_r) = P_{\text{idle}}(c_i) / R_{ir}\), where \(R_{ir}\) denotes the total number of VM instances in the server. We take $R_{ir} = c_i^{\text{core}} / m_r^{\text{core}}$, the number of VMs of this size that fill the server, so that $P_{\text{idle}}(m_r) = P_{\text{idle}}(c_i)\, m_r^{\text{core}} / c_i^{\text{core}}$; cores not used by our VMs are regarded as serving other tenants. The operating power consumption of \(m_r\) is \(P_{\text{run}}(m_r) = P_{\text{max}}(c_i) * \text{CPU\_Util}(m_r) * \frac{m_r^{\text{core}} * m_r^{\text{pflops}}}{c_i^{\text{core}} * c_i^{\text{pflops}}}\). 

We assume that the VM processes all the layer tasks assigned to it based on the space-sharing mechanism. Thus, the computational energy consumption corresponding to layer \(l_k\) allocated to VM instance \(m_r\) is given by 

\begin{equation}
\begin{split}
E_{\text{comp}}(l_k) &= \left[ P_{\text{idle}}(m_r) + P_{\text{run}}(m_r) \right] * T_{\text{exec}}(l_k) \\
&= P_{\text{idle}}(m_r) * T_{\text{exec}}(l_k) + P_{\text{max}}(c_i) * \\
&\frac{X_{l_k}}{c_i^{\text{core}} *c_i^{\text{pflops}}} .
\end{split}
\end{equation}

 where $\text{CPU\_Util}(m_r) = X_{l_k} / (T_{\text{exec}}(l_k) * m_r^{\text{core}} * m_r^{\text{pflops}})$, so the dynamic term does not depend on $T_{\text{exec}}(l_k)$. In the simulator, the static term is charged only for the time the layer occupies the VM, from obtaining the CPU to the end of its computation; waiting in queues and sending the output are not charged, because the delay label already accounts for them.

The transmission energy of a layer is charged per bit: $E_{\text{trans}}(l_k) = \sum_{l_j} 8 D_{l_j} e_{\text{bit}}$ over the direct predecessors $l_j$ placed on other nodes (for the first layer, the upload from the end device; for the last layer, also the return of the result), with $e_{\text{bit}} = 30$~nJ/bit when the data cross the cross-cloud link, which is the 0.03~W per Mb/s of a fixed core network~\cite{mytton2024network}, and 0.2~nJ/bit otherwise, which is the rated power of a data-center switch divided by its port capacity~\cite{cisco2026nexus}. This is $E_{\text{comm}} = D/\text{bw} \cdot (P_{tx} + P_{rx})$ with $(P_{tx} + P_{rx})/\text{bw}$ expressed as the marginal energy per bit. Therefore, for each inference request, the total energy consumption can be expressed as follows.

\begin{equation}
\begin{split}
E_{\text{req}} = \sum_{l_k \in L} \left( E_{\text{comp}}(l_k) + E_{\text{trans}}(l_k) \right).
\end{split}
\end{equation}

\begin{figure*}[t]
\centerline{\includegraphics[width=\textwidth]{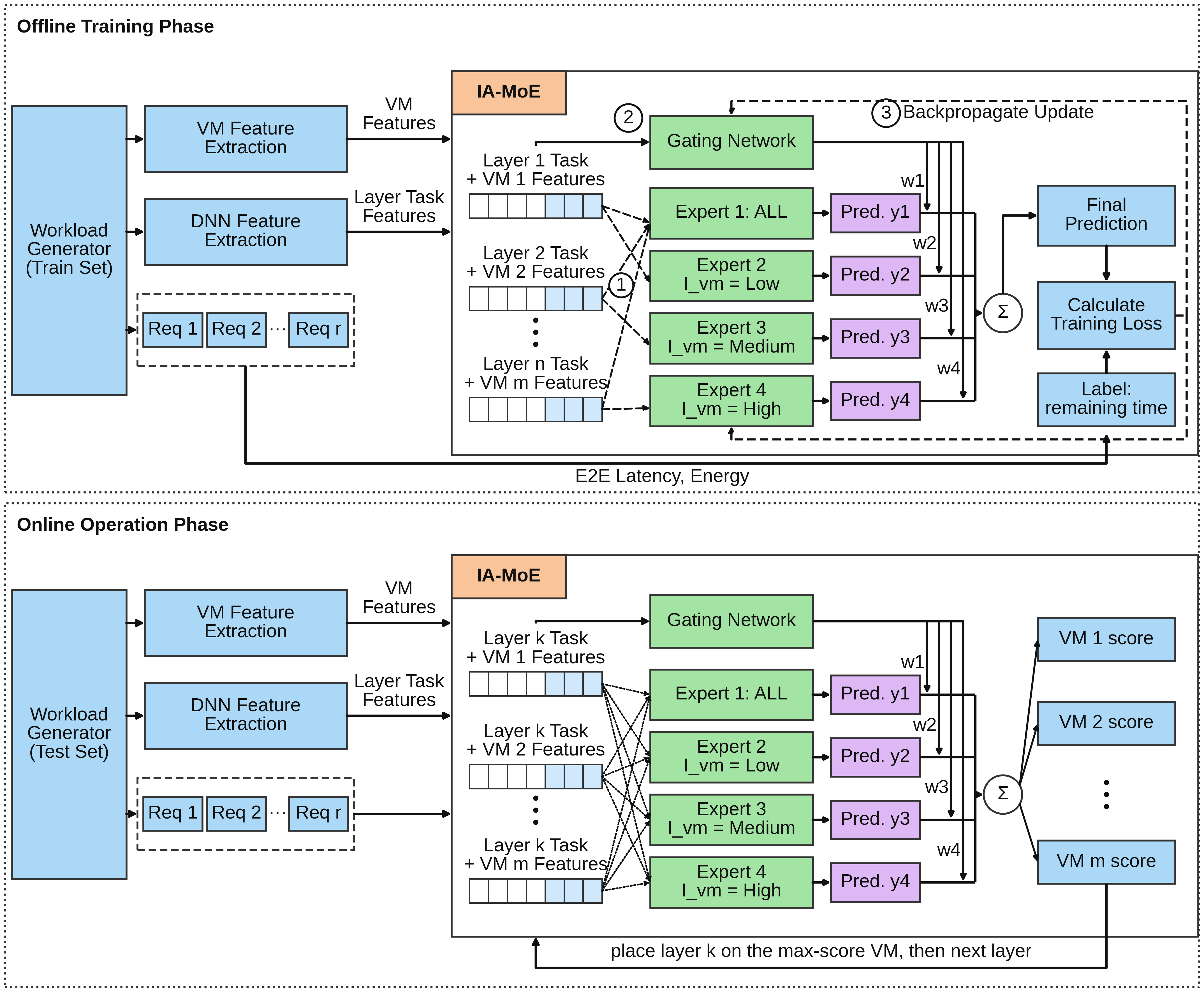}}
\caption{Overview of IA-MoE. Circled numbers mark the training stages: (1) each tier expert is trained on the samples of its intensity tier, and the universal expert (ALL) on all samples; (2) the gating network is trained with the experts frozen; (3) experts and gating are fine-tuned jointly. The label is the remaining-time score computed from the collected runs. Five models trained with different seeds form an ensemble, and their scores are averaged online.}
\label{fig1}
\end{figure*}

\subsection{Optimization Problem}
The problem is formulated using a MIP approach where E2E latency and energy consumption are considered as two objectives. We jointly optimize them for each DNN inference request, subject to constraints \eqref{eq_1}--\eqref{eq_2}. Following the methodology in \cite{lim2024cutting}, we define the optimization target as minimizing the total system cost, \(\text{Cost}_{req}\), which is expressed as a weighted sum of the total energy consumption and E2E latency for each inference request:

\begin{equation}
\begin{split}
\text{Cost}_{req} = \alpha *  T_{req} + \beta * E_{req}.
\end{split}
\label{eq_cost}
\end{equation}

Here, \(\alpha\) and \(\beta\) weight the E2E latency and the energy consumption in the total cost. Accordingly, the MIP problem is formulated as follows: 

$$\text{Minimize} \quad \text{Cost}_{req}$$
$$\text{s.t.} \quad \text{\eqref{eq_1} - \eqref{eq_2}}$$

Since the two terms in the  objective function use different units, they are normalized in the training label (Section~\ref{sec_matching}) and combined by the relative cost RC in the evaluation (Section~\ref{sec_results}).

\section{Methodology}

\subsection{Overview of IA-MoE}
MoE is a learning paradigm where multiple experts specialize in different tasks, with a gating network dynamically weighting each expert by input data and combining their outputs \cite{jacobs1991adaptive}. We extend this idea to DNN partitioning and collaborative inference by formulating layer-to-VM assignment as a matching problem. Each non-virtual layer task can be scheduled to an edge or cloud VM, enabling collaborative execution across heterogeneous resources. Within one request, at most one VM of each server is used; this selection state is reset for every request.

Following the Roofline model \cite{williams2009roofline}, we adopt operational intensity (I), which is the number of multiply-accumulate operations (MACs) per byte of memory exchanged, to categorize VMs into low, medium, and high levels using two thresholds of $I_{low}$ and $I_{high}$.  With $I_{low} = 5$ and $I_{high} = 10$ (MACs per byte), the ten VMs of our setup fall into groups of 3, 3 and 4. Section~\ref{sec_ablation} shows that this grouping has no measurable effect on the delay in our configuration.

To enable efficient model partitioning, we propose the IA-MoE framework, whose core idea is illustrated in Fig.~\ref{fig1}. During the offline training phase, we collect historical information of the current system in advance, including information about DNN models, layer tasks, and the usage of VM resources assigned to layer tasks (Section~\ref{sec_matching}). We use these historical data to construct feature vectors for both layer tasks and VM resources, which are used for matching training in the IA-MoE framework. In the online inference phase, for each load request $r_w \in R$, each layer task is paired with candidate VM resources. Each expert performs forward propagation based on the input combination features and then outputs a predicted matching layer score. The gating network combines these predictions into one score, and the layer is placed on the VM with the highest score. 

\subsection{Feature Extraction}
The DNN model is decomposed into layer tasks, each representing the resource requirements of a network layer. The layer-task features fall into five categories and the VM features into two (25 dimensions in total):

\begin{itemize}
    \item Layer Basic Information: Includes the computational overhead, memory access volume, and intermediate activations of the neural network layers. Since variations in the initial input data sizes can lead to different outcomes, we consider three representative input dimensions - small (64×64), medium (224×224), and large (512×512) with three default input channels. Layer data is obtained via static model analysis.
\end{itemize}

\begin{itemize}
    \item DAG Structure Context: Includes the number of direct predecessor/successor layers and the layer's depth in the DAG topology.
\end{itemize}

\begin{itemize}
    \item Specific Input Dimension: Three input data sizes are encoded using One-Hot encoding.
\end{itemize}

\begin{itemize}
    \item Batch Size: The total input size (e.g., number of images, data length) of each request corresponds to different batch sizes for varying dimensions, affecting actual resource consumption. Static structural features such as computation overhead or memory access volume under a specific input dimension represent per-sample costs, whereas the batch size scales these values to reflect the real workload size.
\end{itemize}

\begin{itemize}
    \item Statistics of VM Types Allocated to Direct Predecessor Layers: When applying MoE to hierarchical task scheduling, state propagation of predecessor dependencies must be addressed. To keep the model small, allocation results are recorded after the placement of each layer is chosen and used as one of the input features for the next layer. Three counts are used: direct predecessors on other edge VMs, on other cloud VMs, and on the candidate VM itself.
\end{itemize}

VM resources refer to a set of resource configurations allocated to a DNN layer, with features describing the upper limits of available resources.

\begin{itemize}
    \item VM Resource Configuration: Includes CPU computing capacity, memory bandwidth, and network bandwidth. For the cloud VM, the network bandwidth is the currently available bandwidth, the smaller of its network interface and the estimated cross-cloud bandwidth.
\end{itemize}

\begin{itemize}
    \item Runtime Resource Utilization: Includes CPU utilization, memory bandwidth utilization, and network bandwidth utilization, reflecting current VM load levels. Because a utilization saturates at 100\% as soon as a VM is busy, the CPU entry is the work already assigned to the VM and not yet finished, in seconds, divided by 10~s, and the network entry is the backlog on the bottleneck link of the path from the predecessors to the candidate VM, divided by the data that the link can send in 10~s. Both are scaled as $\log(1+x)/\log 31$ and capped at 1.
\end{itemize}

\subsection{Network Design}
Expert networks are two-layer fully connected multi-layer perceptrons with an identity function (linear output) in the final layer. Each expert maps the 25 input features to 64 hidden units with ReLU and then to one score, i.e., 1729 parameters.

We design a division of the training data by the Intensity Level corresponding to the VM in the combined features. Specifically, sample data associated with VMs of low, medium, and high intensity train distinct experts.  This division lets each expert network specialize in handling task-resource combinations within its designated category. Additionally, we introduce a universal expert trained on all sample data. Section~\ref{sec_ablation} shows that this division has no measurable effect on the delay in our configuration.

The gating network is a single linear layer followed by Softmax, as in the multi-gate MoE of \cite{ma2018modeling}; it maps the 25 features to the weights of the four experts (104 parameters). The final score is the weighted sum of the expert outputs. Neither the experts nor the gating network use dropout.

\subsection{Matching Training}\label{sec_matching}
Training needs a label for each (layer, VM) sample that reflects the request-level objective. We first define a per-layer score, then show that it makes the greedy placement myopic, and finally define the remaining-time label that we use. The per-layer score is

\begin{equation}
\begin{split}
\mathrm{Score}_{l_k} = \alpha\,\bigl[s(T_{\text{recv}}(l_k), T^{*}(l_k)) + s(T_{\text{req}}, T_{\text{ref}})\bigr] \\
\qquad {}+ \beta\,\bigl[\text{Norm}(1/\log(1+\text{E}(l_k))) + 1 - \text{E}(l_k)/E_{\text{req}}\bigr],
\end{split}
\label{eq_layer_score}
\end{equation}

where \(T_{\text{req}}\) represents the execution time of the entire request; \(\text{E}(l_k)\) represents the inference energy consumption of layer task \(l_k\), and \(E_{\text{req}}\) represents the inference energy consumption of the entire request. $\text{E}$ is in joules; adding one keeps the argument of the logarithm above one, which removes the singularity at 1 and makes the transformation monotonic. The time terms use the slowdown score
\begin{equation}
s(t, t^{*}) = \min\left(1, \max\left(0, 1 - \frac{\ln(t/t^{*})}{\ln(T_{\text{out}}/t^{*})}\right)\right),
\label{eq_slowdown}
\end{equation}
which is 1 when $t$ reaches the reference $t^{*}$ and 0 at the timeout $T_{\text{out}} = 300$~s. $T_{\text{recv}}(l_k)$ is measured at the receiver: from the latest computation end among the direct predecessors of $l_k$ to the computation end of $l_k$, so it includes the transfer of the inputs of $l_k$, which the placement of $l_k$ decides. $T^{*}(l_k)$ is the processing time of $l_k$ on the fastest VM, and $T_{\text{ref}}$ is the shortest E2E delay of a single request of the same type and batch size, measured in isolation (the same reference defines the SLO, Section~\ref{sec_settings}). The references are fixed, so the time terms do not depend on the extreme samples of a data collection. For a failed request, the request term and the energy term are 0, and the layer term is computed for the layers that finished.

The score is designed such that higher values indicate better performance.  The energy terms use the min-max normalized logarithmic reciprocal $\text{Norm}(1/\log(1+\text{E}(l_k)))$ and the complementary share $1 - \text{E}(l_k)/E_{\text{req}}$, which rewards a layer that takes a smaller part of the request energy.

\textit{Myopia of a per-layer score.} At run time, IA-MoE places the layers one by one, each on the VM with the highest predicted score. With Eq.~\eqref{eq_layer_score}, the label of a layer reflects mainly its own delay and energy. Offloading a request to the cloud makes the first offloaded layer slow, because its input crosses the wide-area link, while the gain (faster cloud processing, no further transfers) is recorded in the later layers. A greedy policy trained on such labels therefore keeps almost every request at the edge, even when offloading is better for the request. On Azure-H VGG16 with a mean cross-cloud bandwidth of 600~Mbps, the model trained with Eq.~\eqref{eq_layer_score} offloaded 0.5\% of the layers and had a mean delay of 2.35~s, while HiDP, which mostly offloads requests as a whole, reached 1.04~s. Yet the labels in its training data favored the cloud: for an idle cloud VM against an idle edge VM at the same batch size, they were higher for the cloud in every layer except the first and the last. We therefore train on a label that includes the consequences of the decision:
\begin{equation}
\begin{split}
\mathrm{Score}^{\text{rem}}_{l_k} ={}& 2\alpha\, s\bigl(T_{\text{rem}}(l_k), T^{*}_{\text{rem}}(l_k)\bigr) \\
&+ \frac{\beta}{|L_k|} \sum_{l_j \in L_k} \bigl[\text{Norm}(1/\log(1+\text{E}(l_j))) \\
&\qquad + 1 - \text{E}(l_j)/E_{\text{req}}\bigr],
\end{split}
\label{eq_newrem}
\end{equation}
where $L_k$ is the set of non-virtual layers from $l_k$ to the last one, $T_{\text{rem}}(l_k)$ is the time from the latest computation end among the predecessors of $l_k$ to the end of the request, and $T^{*}_{\text{rem}}(l_k) = \sum_{l_j \in L_k} T^{*}(l_j)$. For a failed request, the time term is 0. The factor 2 gives the time term the same range as the two time terms of Eq.~\eqref{eq_layer_score}. The label is computed from the data collected by the collection policies, so a model trained on it approximates a one-step policy improvement over them \cite{brandfonbrener2021offline}; the placement is still made layer by layer, and only the label changes.

\textit{Data collection.} The training data are collected by running the training workload once with a mixture of policies chosen per request: Random (probability 0.2), HiDP (0.4) and Edge-EFT (0.4) (Section~\ref{sec_baselines}). Every (layer, VM) decision of a request becomes one sample with the label above, including the decisions of failed requests.

\textit{Training.} Training has three stages. First, each expert is trained on its own data subset with the mean squared error (MSE) for 50 epochs. Second, the experts are frozen and the gating network is trained for one pass with the loss $\text{MSE} - \lambda_H H(w)$, where $H(w)$ is the entropy of the expert weights and $\lambda_H = 0.01$; the entropy term penalizes concentrating the weight on one expert. Third, the experts and the gating network are fine-tuned jointly with the same loss for 100 epochs, and the weights of the last 25\% of these epochs are averaged (stochastic weight averaging \cite{izmailov2018averaging}). We train an ensemble of five such models with different seeds and average their scores, which reduces the variation of the placements between seeds. The regression accuracy on the label does not decide the quality of the placements, so we evaluate policies only by simulation.

\section{Experimental Evaluation}
\subsection{Dataset Preprocessing}
We conduct experiments using two publicly available user request workloads datasets. The datasets are split using time-series sampling into chronological windows, with 70\% of the data in each window for training and 30\% for testing (Alibaba: 588 and 250 requests; Azure: 2351 and 1005 requests). 

\begin{figure}[htbp]
\centerline{\includegraphics[width=\linewidth]{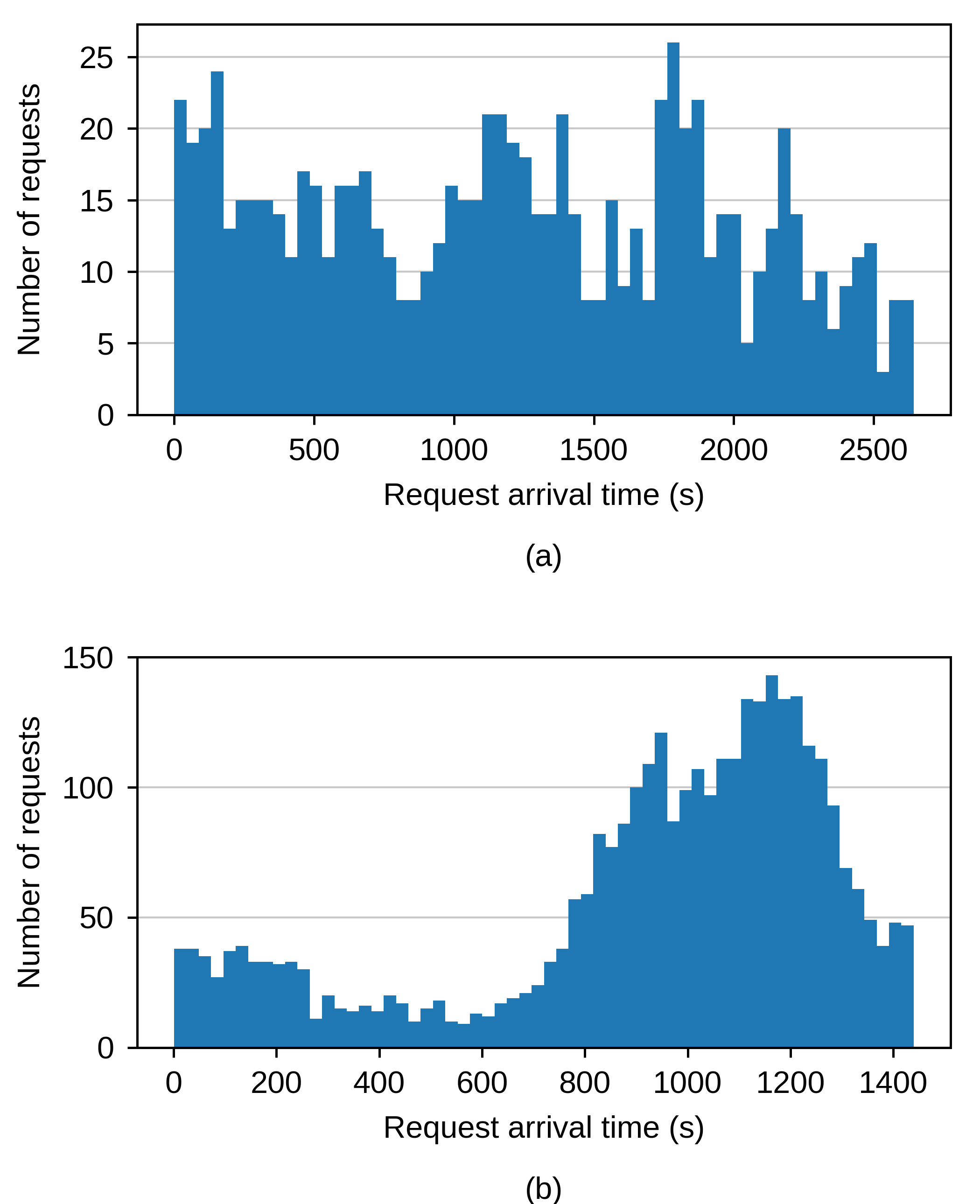}}
\caption{Request arrival distribution (training and test sets, 60 bins). (a) Alibaba. (b) Azure; Azure-H and Azure-S use the same arrivals.}
\label{fig_arrivals}
\end{figure}

\textit{1) Alibaba cluster-trace-gpu-v2025 \cite{yang2025gpu}:} The dataset includes a total of 23871 inference workloads from 156 inference services. We focused on inference instances that only use CPU nodes, and grouped them according to CPU requirements and memory requirements. Three specifications of instances (i.e., $app_0$, $app_{87}$ and $app_{19}$) were selected to simulate the three specifications of the original input data. Each record is the creation of a service instance. After sorting the records by creation time (after time 34340 of the trace), we keep every record that is at least 1000 time units after the previously kept one, which yields 838 records, and divide the creation times by 1000 and subtract 34 to obtain arrival times of 0--2642~s. The histogram of the sample data is shown in Fig.~\ref{fig_arrivals}(a). The number of requests fluctuates between 10 and 25 during most periods, with no obvious long-term peaks or troughs.

\textit{2) Azure LLM inference trace 2024 \cite{stojkovic2025dynamollm}:} This dataset provides tracing samples of multiple LLM inference services from May 10 to 16, 2024 (we use the code trace). We selected the complete data of May 15, 2024, and sampled it every 300 data points. Then, we converted the ContextTokens and GeneratedTokens of all sample data into large, medium, and small categories at their 33\% and 66\% quantiles, and kept only the requests where the category labels of ContextTokens and GeneratedTokens are the same (i.e., both large, both medium, or both small); the category of ContextTokens gives the request type. Finally, 3356 records were retained. The time of day in seconds is divided by 60, so that one day spans 1440~s. The histogram of the sample data and the statistics of requests per second are shown in Fig.~\ref{fig_arrivals}(b). The request distribution exhibits significant diurnal fluctuations and peak characteristics.

\begin{figure}[htbp]
\centerline{\includegraphics[width=\linewidth]{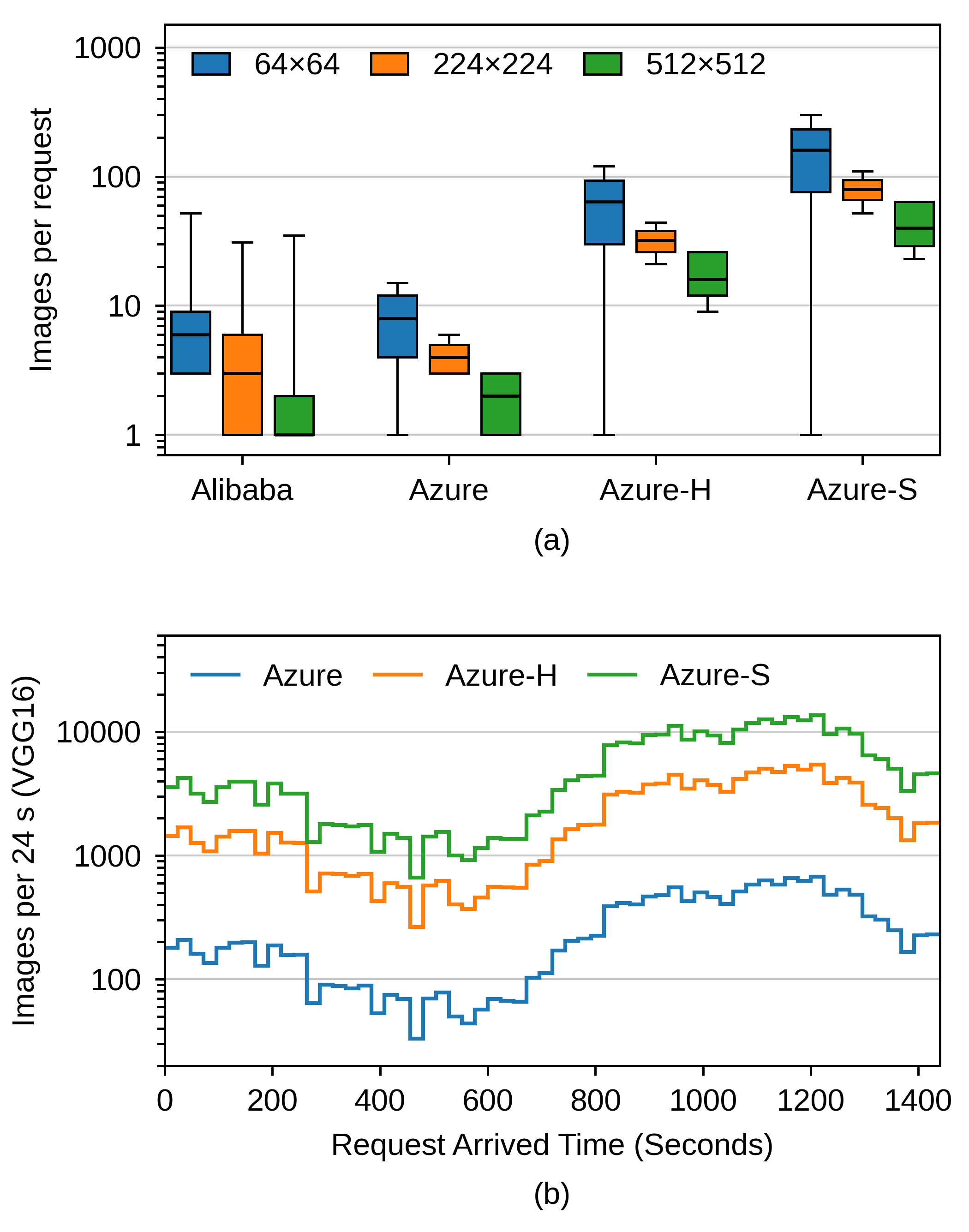}}
\caption{Request sizes and load (training and test sets). (a) Images per request by input size; boxes show the quartiles and the median, whiskers the minimum and maximum. (b) Images per 24~s for VGG16 on Azure, Azure-H and Azure-S.}
\label{fig_request_sizes}
\end{figure}

\textit{3) Request sizes and scenarios:} Each request carries a batch of inputs. The batch is a base value per request type (small, medium and large inputs) times the relative size of the request in its own trace, whose mean is about 1: for Azure, the ContextTokens divided by the median of its class; for Alibaba, which records no request size, the number of instances created in the same window of 1000 time units divided by the mean of its type. The bases are 8/4/2 images for Alibaba and Azure. To create contention, we derive two further workloads from the Azure trace with the same arrival times, types and relative sizes: Azure-H with bases 64/32/16 (high load), and Azure-S with bases 160/80/40 (saturated edge, VGG16 only). The test sets contain 250 (Alibaba) and 1005 (Azure, Azure-H, Azure-S) requests. Fig.~\ref{fig_request_sizes} shows the resulting request sizes and the load over time: the Alibaba sizes have a long tail (up to 52 images), the Azure sizes are bounded, and Azure-H and Azure-S carry about 8 and 20 times the images of Azure at the same arrival times.

\subsection{Experimental Settings}\label{sec_settings}
CloudSimSDN \cite{son2019cloudsimsdn}, an edge computing environment simulator built on CloudSim 4G \cite{beloglazov2012optimal}, models host and network usage in data centers and request response times. Considering that the enhanced virtualization parameters in CloudSim 7G \cite{andreoli2025cloudsim} can improve simulation realism, we upgraded the code of CloudSimSDN to be compatible with this version. We then added terminal device entity to enable user devices to generate request workloads and submit them to edge hosts in data centers. We also extended the task processing logic to support DAG-formatted tasks, dividing tasks into three stages. 

Each layer task occupies all cores of its VM exclusively, so the computing capacity of a VM is proportional to its number of cores, including the cloud VM. The computing capacity per core is 96 G MACs per second (192 GFLOPS), and the network bandwidth is 4 Gbps for edge VMs and 25 Gbps for the cloud VM. In terms of resource configuration, we constructed VM configurations with varying CPU cores (1, 2, 4) and memory bandwidth (16, 32, 64 GB/s). The resulting operational intensity values, shown in Table~\ref{tab2}, fall into three ranges under the thresholds of $I_{low} = 5$ and $I_{high} = 10$. Note that the per-core memory bandwidth of the 1-core VMs with 32 and 64~GB/s and of the 2-core VM with 64~GB/s exceeds what a single core of a current server CPU can sustain (about 19~GB/s measured and 23.6~GB/s in theory on Sapphire Rapids \cite{mccalpin2023bandwidth}); these low-intensity VMs are a simulation setting. For the edge layer, 3 servers are configured, and 3 VMs with different resource combinations are initially deployed on each server; for the cloud layer, 1 VM is initially configured. As for the energy consumption calculation, the idle power of an edge server is 120 watts, and the dynamic power at full load is 154 watts, i.e., 274 watts in total. The cloud server is modeled as a 384-core host (two AMD EPYC 9965) with 129~W at idle and 655~W of dynamic power at full load, from a published SPECpower result; idle power is shared by core count (Section~\ref{sec_settings_energy}).

The edge and the cloud are connected through a cross-cloud link whose bandwidth follows an AR(1) process, $B(t) = \mu + \phi\,(B(t-\Delta) - \mu) + \sigma\sqrt{1-\phi^2}\,\varepsilon$ with $\varepsilon \sim N(0,1)$, $\Delta = 10$~s, $\phi = e^{-\Delta/\tau}$, $\tau = 60$~s, a floor of $0.1\mu$, and $\sigma = \text{CV}\cdot\mu$. The default is $\mu = 150$~Mbps and $\text{CV} = 0.3$ (Azure-S: $\mu = 400$~Mbps). All policies see the same bandwidth trace. The policies and the features do not see the true bandwidth; they use an estimate probed every 10~s. Within a data center, links have a delay of 1~ms; the cross-cloud switch adds 100~ms. In total, the one-way delay from an end device to the cloud VM is 231~ms.

Each request has a timeout of 300~s and is counted as failed if it does not finish within it. The service-level objective (SLO) of a request is $3 T_{\text{ref}}$, where $T_{\text{ref}}$ is the shortest E2E delay of a single request of the same type and batch size, measured on each of the ten VMs in isolation with constant cross-cloud bandwidth $\mu$.

\begin{table}[htbp]
\centering
\caption{Detailed configurations of VM Resources}
\begin{tabular}{|p{1cm}|p{1cm}|p{2cm}|p{2cm}|} \hline 
\textbf{Belonged Server} & \textbf{CPU Cores} & \textbf{Memory Bandwidth (GB/s)} & \textbf{Operational Intensity (MACs/Byte)} \\ \hline 
\text{Edge 1} & 1 & 16 & 5.5879 \\ \hline
\text{Edge 1} & 1 & 32 & 2.7940 \\ \hline
\text{Edge 1} & 1 & 64 & 1.3970 \\ \hline
\text{Edge 2} & 2 & 16 & 11.1759 \\ \hline
\text{Edge 2} & 2 & 32 & 5.5879 \\ \hline
\text{Edge 2} & 2 & 64 & 2.7940 \\ \hline
\text{Edge 3} & 4 & 16 & 22.3517 \\ \hline
\text{Edge 3} & 4 & 32 & 11.1759 \\ \hline
\text{Edge 3} & 4 & 64 & 5.5879 \\ \hline
\text{Cloud} & 64 & 256 & 22.3517 \\ \hline
\end{tabular}
\label{tab2}
\end{table}

The DNN models adopted in the experiments include AlexNet, VGG16, and ResNet50. With the help of the open-source tool torchinfo, we obtained the number of parameters of layers, computing overhead, and the size of intermediate output data.  The training hyperparameters (epochs, batch size, learning rate, weight decay, hidden width, learning-rate schedule, weight averaging and $\lambda_H$) were chosen by a random search of 30 configurations, scored on a held-out 15\% of the collected requests, and a simulation of the best candidates; RF uses the default settings of Weka. Specifically, the main parameters used by our proposed approach are shown in Table~\ref{tab3}.

\begin{table}[htbp]
\centering
\caption{Hyper Parameters of IA-MoE}
\begin{tabular}{|c|c|} \hline
\textbf{Parameter} & \textbf{Value} \\ \hline
\text{Expert hidden units} & 64 \\ \hline
\text{Gating hidden units} & none (linear + Softmax) \\ \hline
\text{Learning rate} & 0.0014 \\ \hline
\text{Batch size} & 64 \\ \hline
\text{Optimizer} & AdamW \\ \hline
\text{Training epochs} & 50 (experts), 100 (joint) \\ \hline
\text{Weight decay} & 0.01 \\ \hline
\text{Dropout rate} & 0 \\ \hline
Entropy weight $\lambda_H$ & 0.01 \\ \hline
Weight averaging & last 25\% of joint epochs \\ \hline
Ensemble size & 5 \\ \hline
\end{tabular}
\label{tab3}
\end{table}

\subsection{Baselines}\label{sec_baselines}
\begin{itemize}
    \item HiDP \cite{taufique2024hidp}: Uses a two-level (global-local) hierarchical partitioning approach with dynamic programming (DP) to determine optimal partitioning modes, calculating total computation time for model and data partitioning to select the faster strategy. Data partitioning is not modeled, because a layer task occupies a whole VM in the simulator. It plans each request from the static delay model and does not see the current load of the VMs or links.
\end{itemize}

\begin{itemize}
    \item RF \cite{chen2021dnnoff}: Investigates offloading DNN-based IoT applications in the end-edge-cloud architecture. For a given DNN application, DNNOff first rewrites source code to support on-demand offloading, then uses a random forest model to predict per-layer execution delays. We extracted the random forest model to predict weighted target scores specified in this study. It is trained on the same data and labels as IA-MoE, with the Weka defaults (100 trees, unlimited depth).
\end{itemize}

\begin{itemize}
    \item Random: At the start of a request, one random VM is drawn per server, and the request uses only these VMs. A request uses the cloud with probability 0.24. The first layer goes to a random one of the allowed VMs; from the second layer on, each layer switches to another of them with probability 0.25, and the cloud is used for at most one contiguous segment.
\end{itemize}

\begin{itemize}
    \item Edge-EFT: Places the whole request on the edge VM with the earliest estimated finish time, computed from the queued work of the VM and the processing time of the request, in the spirit of HEFT-style list scheduling. It never uses the cloud.
\end{itemize}

\begin{itemize}
    \item Double DQN \cite{vanhasselt2016deep}: An online reinforcement learning agent that makes the same layer-by-layer decisions with the same 25 features. Its Q network has the size of one IA-MoE expert (25-64-1). The reward of a decision is the per-layer part of Eq.~\eqref{eq_layer_score} divided by the number of layers, plus $\alpha\, s(T_{\text{req}}, T_{\text{ref}})$ at the last layer, with no discount within a request, so the return covers the whole request. An episode is one simulation of the training workload; after each episode, the agent performs one update of batch 64 per four decisions from a replay buffer of 200{,}000 transitions, and synchronizes the target network every 10{,}000 updates. Exploration is $\epsilon$-greedy with $\epsilon$ decaying linearly from 1 to 0.05 over 20 episodes, either per decision (step) or with temporally extended actions ($\epsilon z$-greedy, segment) \cite{dabney2021temporally}. Each run trains for 120 episodes; every 5 episodes the network is evaluated greedily on the training workload, and the best checkpoint is kept, as in \cite{vanhasselt2016deep}. The optimizer settings equal those of IA-MoE.
\end{itemize}

\subsection{Experimental Results}\label{sec_results}

\textit{Metrics.} For each method we report the mean E2E delay over all requests, where a failed request counts as 300~s, the mean energy of the completed requests, and the share of layers placed in the cloud. Learned methods and Random are run with five seeds (42--46); we report the mean and, where relevant, the range over seeds. We call a difference meaningful only if the means differ by at least 3\% and the seed ranges do not overlap; this rule was fixed before the ablation and DQN runs. To combine delay and energy, we use the relative cost $\text{RC} = (\bar{T}/\bar{T}_{\text{HiDP}})^{\alpha} (\bar{E}/\bar{E}_{\text{HiDP}})^{\beta}$ with the weights of Eq.~\eqref{eq_cost} ($\alpha = 0.6$, $\beta = 0.4$), and its geometric mean over scenarios \cite{fleming1986not}. RC does not depend on the choice of reference: another reference multiplies all values by one constant and leaves the ranking unchanged. A Z-score over the compared methods lacks this property: one method that fails many requests compresses the differences between the others.

\begin{table*}[t]
\centering
\caption{Mean E2E delay (s, failed requests count as 300~s) / mean energy per completed request (J). Learned methods and Random: mean of five seeds. RC: relative cost to HiDP, geometric mean over the nine non-saturated scenarios.}
\label{tab_main}
\begin{tabular}{|l|c|c|c|c|c|} \hline
\textbf{Scenario} & \textbf{IA-MoE} & \textbf{RF} & \textbf{Edge-EFT} & \textbf{HiDP} & \textbf{Random} \\ \hline
Alibaba AlexNet & 0.05 / 0.34 & 0.06 / 0.38 & 0.05 / 0.35 & 0.05 / 0.35 & 0.21 / 0.64 \\ \hline
Alibaba VGG16 & 0.30 / 4.87 & 0.28 / 4.95 & 0.25 / 4.87 & 0.24 / 4.22 & 1.22 / 9.79 \\ \hline
Alibaba ResNet50 & 0.10 / 1.42 & 0.10 / 1.42 & 0.10 / 1.43 & 0.10 / 1.42 & 0.97 / 4.53 \\ \hline
Azure AlexNet & 0.05 / 0.35 & 0.05 / 0.34 & 0.05 / 0.35 & 0.05 / 0.35 & 0.22 / 0.68 \\ \hline
Azure VGG16 & 0.27 / 5.01 & 0.27 / 5.02 & 0.26 / 5.03 & 0.32 / 5.00 & 1.47 / 10.33 \\ \hline
Azure ResNet50 & 0.11 / 1.46 & 0.11 / 1.47 & 0.10 / 1.47 & 0.11 / 1.46 & 1.02 / 4.28 \\ \hline
Azure-H AlexNet & 0.27 / 2.04 & 0.27 / 2.04 & 0.27 / 2.04 & 0.28 / 2.03 & 0.95 / 4.70 \\ \hline
Azure-H VGG16 & \textbf{2.10} / 33.49 & 2.36 / 36.50 & 2.46 / 44.54 & 42.51 / 27.10 & 61.33 / 73.59 \\ \hline
Azure-H ResNet50 & 0.85 / 11.85 & 0.87 / 11.79 & \textbf{0.79} / 11.93 & 2.14 / 11.53 & 37.60 / 29.50 \\ \hline
RC (9 scenarios) & 0.784 & 0.802 & 0.785 & 1.000 & 3.884 \\ \hline
Azure-S VGG16 & \textbf{5.13} / 69.23 & 5.41 / 72.24 & 52.65 / 135.59 & 99.59 / 52.55 & 192.13 / 167.02 \\ \hline
RC (Azure-S) & 0.188 & 0.198 & 0.997 & 1.000 & 2.356 \\ \hline
\end{tabular}
\end{table*}

\textit{1) Impact of Dynamic Workloads:}  Table~\ref{tab_main} shows the results. On the light workloads (Alibaba, Azure, and Azure-H with AlexNet), the learned methods and Edge-EFT place no layer in the cloud, the edge is lightly used, and all methods except Random reach nearly the same delay and energy; the largest gap is on Alibaba VGG16, where IA-MoE is slower than HiDP (0.30 vs.\ 0.24~s). On Azure-H ResNet50, Edge-EFT is the fastest (0.79 vs.\ 0.85~s for IA-MoE). Over the nine non-saturated scenarios, IA-MoE and Edge-EFT have the same RC (0.784 and 0.785). The learned methods gain over the heuristics only where the edge is congested and offloading pays off: on Azure-H VGG16, IA-MoE has the lowest delay (2.10~s against 2.36~s for RF and 2.46~s for Edge-EFT), and on Azure-S, where the edge alone cannot keep up, it is 10 times faster than Edge-EFT and 19 times faster than HiDP, and 5\% faster than RF (5.13 vs.\ 5.41~s). HiDP plans without the current load and sends too many requests over the congested cross-cloud link.

\textit{2) Removing the Myopia of the Per-Layer Score:} Table~\ref{tab_label} compares training on the per-layer score of Eq.~\eqref{eq_layer_score} with training on the remaining-time label of Eq.~\eqref{eq_newrem}. With the per-layer score, the learned policies hardly use the cloud even when the cross-cloud bandwidth is high. With the remaining-time label, IA-MoE offloads 15--64\% of the layers in these scenarios and reduces the mean delay by 12\% to 56\%; at $\mu = 400$ and 600~Mbps it matches or beats HiDP, while the per-layer model stays at 2.35--2.37~s. RF improves in the same direction; on Azure-H it gains less than IA-MoE, and on Azure-S slightly more (14\% vs.\ 12\%). On the light workloads the label changes little; on Alibaba VGG16 the delay of IA-MoE rises from 0.25 to 0.30~s.

\begin{table}[htbp]
\centering
\caption{Per-layer score vs.\ remaining-time label: mean E2E delay (s) and cloud share of layers (in parentheses), VGG16, mean of five seeds.}
\label{tab_label}
\begin{tabular}{|l|c|c|c|} \hline
\textbf{Scenario} & \textbf{Per-layer} & \textbf{Remaining} & \textbf{HiDP} \\ \hline
Azure-H, $\mu$ = 150 & 2.43 (0.00) & 2.10 (0.15) & 42.51 \\ \hline
Azure-H, $\mu$ = 400 & 2.37 (0.00) & 1.28 (0.58) & 1.48 \\ \hline
Azure-H, $\mu$ = 600 & 2.35 (0.00) & 1.03 (0.64) & 1.04 \\ \hline
Azure-S, $\mu$ = 400 & 5.80 (0.15) & 5.13 (0.46) & 99.59 \\ \hline
\end{tabular}
\end{table}

\begin{figure*}[t]
\centerline{\includegraphics[width=\textwidth]{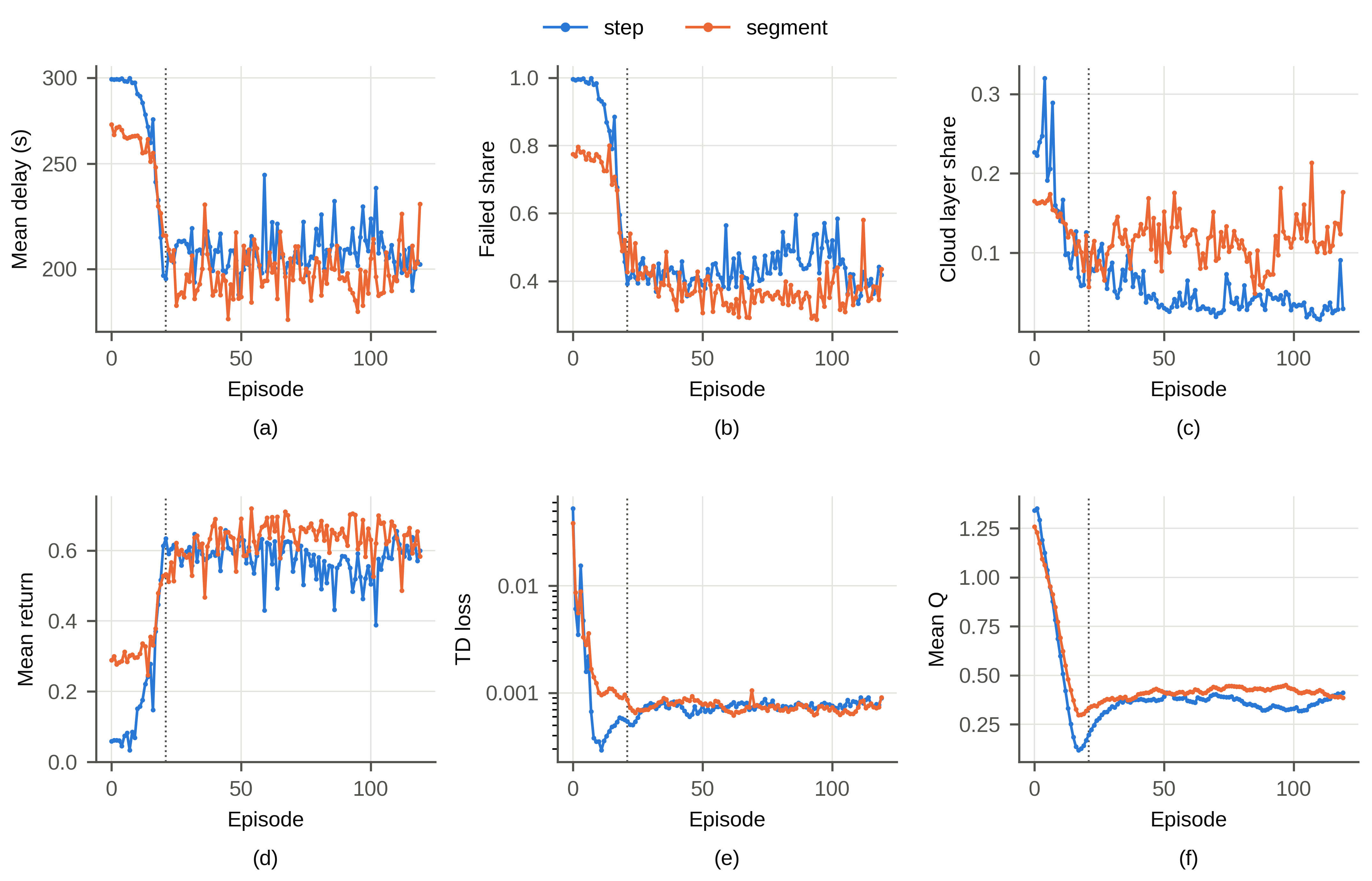}}
\caption{Double DQN learning curves on the Azure-S VGG16 training workload (seed 42), per episode, for step and segment exploration: (a) mean E2E delay on the training workload (failed requests count as 300~s); (b) share of failed requests; (c) share of layers placed in the cloud; (d) mean return per request; (e) mean TD loss per update; (f) mean Q value of the sampled transitions. (a) and (e) use logarithmic axes. The dotted line marks the episode at which $\epsilon$ reaches 0.05.}
\label{fig_dqn}
\end{figure*}

\begin{figure*}[t]
\centerline{\includegraphics[width=\textwidth]{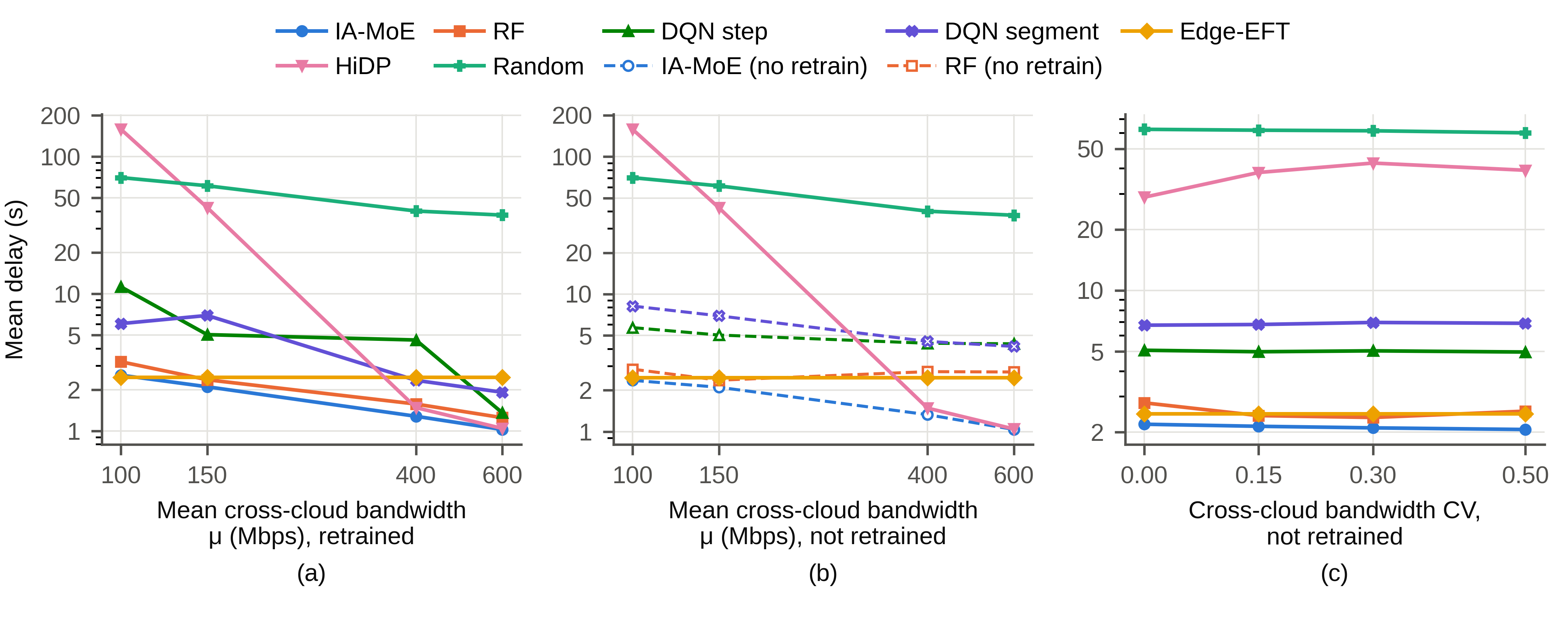}}
\caption{Mean E2E delay on Azure-H VGG16 (log scale; failed requests count as 300~s) as the cross-cloud bandwidth changes. (a) Mean bandwidth $\mu$ with CV = 0.3, learned methods retrained at each $\mu$. (b) The same, learned methods trained at $\mu = 150$~Mbps and not retrained (dashed); Edge-EFT, HiDP and Random need no training and are the same as in (a). (c) Coefficient of variation with $\mu = 150$~Mbps, learned methods trained at CV = 0.3 and not retrained. The learned methods are IA-MoE, RF and Double DQN with step and segment exploration. The $\mu$ axes in (a) and (b) are logarithmic.}
\label{fig_bandwidth}
\end{figure*}

\begin{figure*}[t]
\centerline{\includegraphics[width=\textwidth]{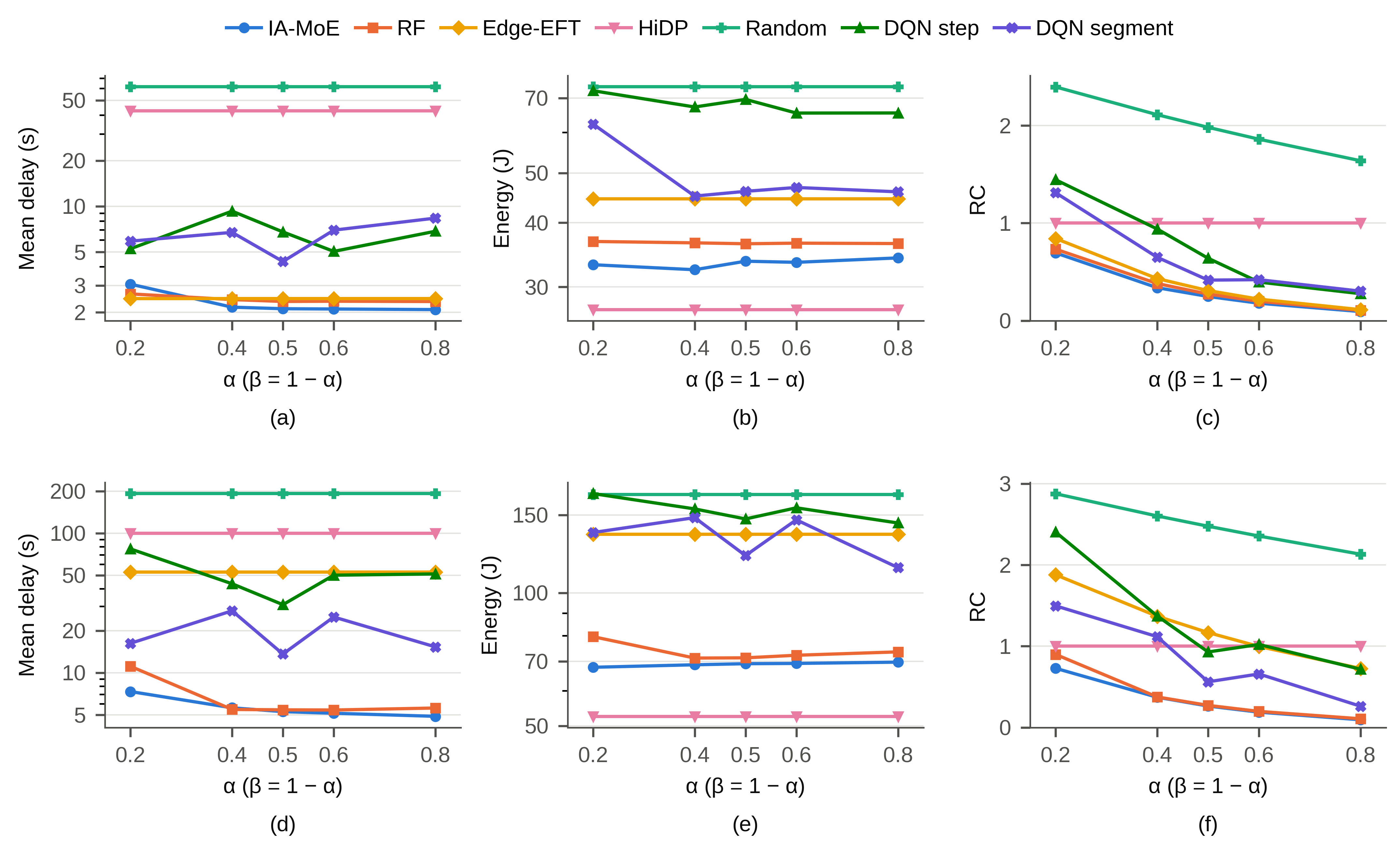}}
\caption{Effect of the time weight $\alpha$ ($\beta = 1 - \alpha$) on VGG16, with the labels regenerated and IA-MoE and RF retrained at each $\alpha$, and Double DQN (step and segment exploration) retrained with the same weights in its reward; top: Azure-H, bottom: Azure-S. (a), (d) Mean E2E delay (failed requests count as 300~s) and (b), (e) mean energy per completed request, both on a log scale; (c), (f) RC with the same $\alpha$ and $\beta$, with HiDP as the reference. The baselines do not depend on $\alpha$.}
\label{fig_alpha}
\end{figure*}

\textit{3) Comparison with Deep Reinforcement Learning:} Table~\ref{tab_dqn} compares IA-MoE with the Double DQN agent on three VGG16 scenarios (for $\mu = 600$~Mbps, the DQN agent is trained on the Azure-H training workload at $\mu = 600$). Both exploration variants are worse than IA-MoE in all three scenarios, with seed ranges that do not overlap; the gap grows with congestion. The same holds in every bandwidth setting of Fig.~\ref{fig_bandwidth} and for every $\alpha$ in Fig.~\ref{fig_alpha}. Fig.~\ref{fig_dqn} points to a likely cause on Azure-S: after $\epsilon$ reaches 0.05, 29--60\% of the training requests still fail (Fig.~\ref{fig_dqn}(b)), so most transitions in the replay buffer come from failed requests, and every change of the policy changes the load that the next episode sees. IA-MoE does not depend on samples from its own policy: its labels come from the logged outcomes of the collection policies.

Imitation learning has been reported to beat DRL for computation offloading when optimal decisions can be computed offline \cite{yu2020intelligent, chen2020knowledge}; here such decisions are not available, and learning from logged outcomes still beats Double DQN. Training also costs much less: one DQN run of 120 episodes took 84--155 minutes (six runs in parallel on a shared machine), while data collection for IA-MoE took 5--19~s, and training its ensemble of five models together with the test run took 20--418~s. The DQN agent treats requests as independent episodes, uses no discount within a request and trains for 120 episodes; we did not test longer training or policy-gradient methods such as PPO.

\begin{table}[htbp]
\centering
\caption{IA-MoE vs.\ Double DQN: mean E2E delay (s), mean [range] over five seeds.}
\label{tab_dqn}
\begin{tabular}{|l|c|c|c|} \hline
\textbf{Method} & \textbf{Azure-H 150} & \textbf{Azure-S} & \textbf{Azure-H 600} \\ \hline
IA-MoE & 2.10 [2.08--2.11] & 5.13 [4.98--5.33] & 1.03 [1.02--1.03] \\ \hline
DQN step & 5.04 [2.88--7.83] & 50.00 [25.9--88.2] & 1.36 [1.27--1.48] \\ \hline
DQN segment & 6.95 [4.57--12.0] & 25.09 [11.4--48.5] & 1.92 [1.29--3.44] \\ \hline
RF & 2.36 & 5.41 & 1.25 \\ \hline
HiDP & 42.51 & 99.59 & 1.04 \\ \hline
\end{tabular}
\end{table}

\textit{4) Robustness:} Fig.~\ref{fig_bandwidth} varies the cross-cloud bandwidth on Azure-H VGG16. When the mean $\mu$ changes (Fig.~\ref{fig_bandwidth}(a)), IA-MoE retrained at each $\mu$ has the lowest mean delay for $\mu \geq 150$~Mbps; at $\mu = 100$~Mbps, where offloading does not pay, it is slightly slower than Edge-EFT (2.55 vs.\ 2.46~s). The model trained at $\mu = 150$ and not retrained performs about the same (2.36, 1.33 and 1.04~s at $\mu$ = 100, 400 and 600; Fig.~\ref{fig_bandwidth}(b)), while the RF model without retraining hardly offloads (1--3\% of the layers). When the variability CV changes from 0 to 0.5 without retraining (Fig.~\ref{fig_bandwidth}(c)), the delay of IA-MoE stays between 2.06 and 2.19~s, below Edge-EFT (2.46~s) and RF. Both Double DQN variants are slower than IA-MoE in every setting of Fig.~\ref{fig_bandwidth}, with non-overlapping seed ranges: retrained at each $\mu$ they take 1.36--11.23~s, and trained at $\mu = 150$ without retraining they take 4.16--8.19~s, more than twice the delay of IA-MoE.

\textit{5) Sensitivity to $\alpha$ and $\beta$:}\label{sec_alpha} We vary the time weight $\alpha \in \{0.2, 0.4, 0.5, 0.6, 0.8\}$ with $\beta = 1 - \alpha$ on Azure-H and Azure-S (VGG16). For each $\alpha$, the labels are regenerated and IA-MoE and RF are retrained; the RC uses the same weights (Fig.~\ref{fig_alpha}). In our simulator, a faster VM usually also uses less energy, so delay and energy rarely conflict in the choice of a VM, and the energy term of the score separates placements only weakly (within a layer and batch, its spread is 6 to 10 times smaller than that of the time term). As $\alpha$ grows from 0.2 to 0.8, the delay of IA-MoE falls (Azure-H: 3.06 to 2.08~s; Azure-S: 7.34 to 4.87~s), while its energy changes little: it stays between 32.4 and 34.2~J on Azure-H and rises from 67.9 to 69.7~J on Azure-S (Fig.~\ref{fig_alpha}(b), (e)). HiDP uses less energy than IA-MoE, but its delay is 14 to 20 times higher. IA-MoE has the lowest RC at every $\alpha$ in both scenarios. At $\alpha = 0.2$ on Azure-H, where energy dominates the label, IA-MoE is slower than RF and Edge-EFT (3.06 vs.\ 2.64 and 2.46~s) but still has the lowest RC. Double DQN is also retrained at each $\alpha$, with the same weights in its reward. Both of its variants are slower than IA-MoE at every $\alpha$ in both scenarios, with non-overlapping seed ranges (Azure-H: 4.32--9.28~s; Azure-S: 13.62--77.12~s), use 1.3 to 2.5 times the energy of IA-MoE, and have a higher RC.

\textit{6) Ablation of Intensity Awareness:}\label{sec_ablation} We test whether the intensity-based division of experts and the MoE structure matter, on the three VGG16 scenarios of Table~\ref{tab_dqn}. \emph{Random tiers} shuffles the intensity labels of the samples before dividing them among the tier experts (tier sizes unchanged). \emph{Single network} trains only one network with the same training schedule and no gating, either with 64 hidden units (one expert) or with 260 hidden units, which has the same number of parameters as the full model (7021 vs.\ 7020). Table~\ref{tab_ablation} shows that no variant differs from the full model by our criterion: all seed ranges overlap with the full model. The single network with 64 units is 6.8\% slower on Azure-S, but with overlapping ranges, and widening it to 260 units halves this gap. On both Azure-H scenarios every variant is faster than RF (2.08--2.10 vs.\ 2.36~s and 1.02--1.03 vs.\ 1.25~s); on Azure-S the variants take 5.05--5.48~s and RF 5.41~s. The gain therefore comes from the label and the neural learner, not from the division of experts.

A Roofline bound explains this result. With the processing-time model $T_{\text{process}}$ of Section~\ref{sec_problem}, we compute, for every request in the test workloads, the gain of placing each layer on its fastest VM over placing the whole request on the single best VM, ignoring transfers; this is an upper bound for any intensity-aware choice. In our configuration the bound is 0 for AlexNet, VGG16 and ResNet50: the 4-core VM with 64~GB/s has both the highest compute and the highest memory bandwidth at the edge, so it is the fastest for every layer. The bound is positive only when one device has more compute and another more memory bandwidth. With realistic per-core parameters, we found this only for an edge site with a T4 GPU and a 28-core HBM CPU: the bound is 0.3--4.9\% on average for VGG16 (19.2\% at most) and 0.2--12.5\% for AlexNet (22.8\% at most), larger on small batches. Such configurations are where intensity awareness could matter; testing them is future work.

\begin{table}[htbp]
\centering
\caption{Ablation: mean E2E delay (s), mean [range] over five seeds; change relative to the full model.}
\label{tab_ablation}
\begin{tabular}{|l|c|c|c|} \hline
\textbf{Variant} & \textbf{Azure-H 150} & \textbf{Azure-S} & \textbf{Azure-H 600} \\ \hline
Full IA-MoE & 2.10 [2.08--2.11] & 5.13 [4.98--5.33] & 1.03 [1.02--1.03] \\ \hline
Random tiers & 2.08, $-$0.9\% & 5.05, $-$1.5\% & 1.02, $-$0.1\% \\ \hline
Single, $h$ = 64 & 2.10, $-$0.2\% & 5.48, +6.8\% & 1.03, +0.1\% \\ \hline
Single, $h$ = 260 & 2.09, $-$0.4\% & 5.28, +3.0\% & 1.03, +0.1\% \\ \hline
RF & 2.36 & 5.41 & 1.25 \\ \hline
\end{tabular}
\end{table}

\section{Conclusions}
We propose IA-MoE, an intensity-aware MoE framework that dynamically determines the partitioning and scheduling of DNN inference tasks by matching each layer task to a VM. To validate our approach, we extend CloudSimSDN to capture DAG tasks and VM heterogeneity. Experiments are conducted on two public inference request patterns, three representative DNN models, and ten VM configurations (nine edge VMs and one cloud VM). The main finding is that greedy layer-by-layer placement needs a label that sees the consequences of a decision: a per-layer score keeps most requests at the edge even when offloading is better, and a remaining-time label removes this myopia. With this label, IA-MoE has the lowest mean delay of all compared methods on the congested and saturated VGG16 workloads for mean cross-cloud bandwidths from 150 to 600~Mbps, and a random forest trained on the same label is 5--27\% slower; at 100~Mbps, offloading does not pay and Edge-EFT is 0.09~s faster. On light workloads, IA-MoE performs on par with simple heuristics such as Edge-EFT, and HiDP, which offloads whole requests, uses less energy at up to 20 times the delay. Trained on the logged outcomes of simple collection policies, IA-MoE also beats a Double DQN agent with the same features in all 19 settings we tested, at a small fraction of the training cost.

A Roofline bound shows that dividing experts by operational intensity can help only when some devices have more compute and others more memory bandwidth; this is not the case in our configuration, and ablations against randomly assigned tiers and against a single network of the same size confirm that the division has no measurable effect there. We did not simulate a configuration with such devices. The gain of the remaining-time label is shown on VGG16; with AlexNet and ResNet50, offloading does not pay in our workloads (Edge-EFT, which never offloads, is as fast as the learned methods or faster), so these models do not test it, and the ablation, the DQN comparison and the bandwidth study use VGG16 only for the same reason. The DQN result holds for Double DQN trained for 120 episodes; longer training and policy-gradient methods were not run. Because faster VMs also use less energy in our simulator, the energy term of the score separates placements only weakly, and the $\alpha$ study changes mainly the delay; it does not test a real delay--energy trade-off.

Future work is to test the intensity-based division on a configuration with a positive Roofline bound, such as an edge site with a T4 GPU and a 28-core HBM CPU.

\bibliographystyle{ieeetr}
\bibliography{reference}

\end{document}